\documentclass[fleqn,twoside]{article}
\usepackage[headings]{espcrc2}

\readRCS
$Id: espcrc2.tex,v 1.2 2004/02/24 11:22:11 spepping Exp $
\usepackage{graphicx, balance}
\usepackage[figuresright]{rotating}

\newcommand{\AmS}{{\protect\the\textfont2
  A\kern-.1667em\lower.5ex\hbox{M}\kern-.125emS}}

\title{\textbf{COMPARISON ANALYSIS IN  MULTICAST AUTHENTICATION BASED ON BATCH SIGNATURE (MABS) IN NETWORK SECURITY}}

\author{Srikanth Bethu\address[DCSE]{Assistant Professor,Department of Computer Science and Engineering, Holymary Institute of Technology and Science, JNTU Hyderabad - 501 301 India,
Contact: srikanthbethu@gmail.com\\},
K Kanthi Kumar\address{Associate Professor, Department of Electronics and Communications Engineering, Holymary Institute of Technology and Science, Hyderabad},MD Asrar Ahmed\address{M.tech ,Department of Computer Science and Engineering, Osmania University,Hyderabad.},S.Soujanya\address{Assistant Professor,Department of Computer Science and Engineering, Holymary Institute of Technology and Science, JNTU Hyderabad}}

\runtitle{Preparation of International Journal of Information Processing Paper in Two-Column Format}
\runauthor{Srikanth Bethu, et al.,}

\begin{document}
\begin{abstract}

Conventional block-based multicast authentication schemes overlook the heterogeneity of receivers by letting the sender choose the block size, divide a multicast stream into blocks, associate each block with a signature, and spread the effect of the signature across all the packets in the block through hash graphs or coding algorithms. The correlation among packets makes them vulnerable to packet loss, which is inherent in the Internet and wireless networks. Moreover, the lack of Denial of Service (DoS) resilience renders most of them vulnerable to packet injection in hostile environments. In this paper, we propose a novel multicast authentication protocol, namely MABS, including two schemes. The basic scheme (MABS-B) eliminates the correlation among packets and thus provides the perfect resilience to packet loss, and it is also efficient in terms of latency, computation, and communication overhead due to an efficient cryptographic primitive called batch signature, which supports the authentication of any number of packets simultaneously.so we discuss their comparisons and performance evaluation of Packet Loss, Comparisons over Lossy Channels, Comparisons of Signature Schemes, computationational overheads etc.  \\\\
{\bf Keywords :} Denial of Service (DoS) ,MABS,MABS-BSignature Schemes.
\end{abstract}

\maketitle

\section{INTRODUCTION}

Generally, there are following issues in real world challenging the design.
Efficiency: While the sender of multimedia content is usually a powerful server, receivers can have different capabilities and resources.
Resilience to packet loss: Packet may be lost during wireless transmission. In the Internet, congestion at routers is a major reason causing packet loss.

Resilience to denial of service (DoS) attacks: Forged packets injected into a multicast stream increase the workload of receivers and cause the drop of authentic packets, leading to DoS. A certain level of resilience to DoS attacks should be provided. Recently, batch signature schemes can be used to improve the performance of broadcast authentication [2], [3].

 In this paper, we present comprehensive study on this approach and propose a novel multicast authentication protocol called MABS. MABS uses an efficient asymmetric cryptographic primitive called batch signature [3],[4], [5], which supports the authentication of any number of packets simultaneously with one signature verification, to address the efficiency and packet loss problems. MABS provides data integrity,data origin authentication, and non-repudiation. In addition, we make the following contributions:

\renewcommand{\labelenumi}{(\roman{enumi})}
\begin{enumerate}

\item Our MABS can achieve perfect resilience to Packet loss in lossy channels in the sense that no matter how many packets are lost the already received packets can still be authenticated by receivers.

\item MABS-B is efficient in terms of less authentication latency, computation, and communication overhead. Though MABS-E is less efficient than MABS-B since it includes the DoS defense.

\item We propose two new batch signature schemes based on BLS and DSA and show they are more efficient than the batch RSA [5] signature scheme.
\end{enumerate}

\subsection{Properties of Multicast:}

The definition of the host group model provides a summary of the key properties of multicast: Òa host group is a set of network entities sharing a common identifying multicast address, all receiving any data packets addressed to this multicast address by senders (sources) that may or may not be members of the same group and have no knowledge of the
groupsÕ membership.Ó This definition highlights the three main properties of multicast:

\renewcommand{\labelenumi}{(\roman{enumi})}
\begin{enumerate}

\item All members receive all packets sent to the address: Multicast routing delivers all packets sent to the multicast address to all members of the multicast group.

\item Open group membership: Multicast provides an open group model and allows group membership to be transparent to the source.

\end{enumerate}

\subsection{Multicast Routing Protocol:}

Multicast routers execute a multicast routing protocol to define delivery paths that enable the forwarding of multicast datagrams across an internetwork. The Distance Vector Multicast Routing Protocol (DVMRP) is a distance-vector routing protocol, and Multicast OSPF (MOSPF) is an extension to the OSPF link-state unicast routing protocol.

A multicast protocol enables a sender to efficiently disseminate digital media data to many receivers. Due to the time sensitive requirement of some applications, reliable transmission protocol like TCP (Transmission Control Protocol) is impractical for multicast. Therefore, unreliable transmission protocol such as UDP (User Datagram Protocol) is generally adopted for multicast applications. Multicast protocol is suitable for many applications, e.g. video transmissions, live broadcasts, stock quotes, or news feeds. These applications may have many receivers or distribute time-sensitive data. To ensure secure communications between a sender and its receivers, it is important to implement security measures in a multicast environment.

\begin{table}
\renewcommand{\baselinestretch}{1}
\caption{Difference between TCP, UDP}
\begin{small}
\begin{center}
\begin{tabular}{|r|r|} \hline
       TCP& UDP  \\ & \\ \hline
 $\rightarrow $ Reliable&Unreliable\\ \hline
 $\rightarrow$ Connection-oriented&Connectionless \\ \hline
 $\rightarrow$ Segment retransmission
  and flow control
  through windowing &No windowing or retransmission \\ \hline
 $\rightarrow$Acknowledge segments&No acknowledgement\\ \hline
\end{tabular}
\end{center}
\end{small}
\end{table}

\subsection{Security Issues and Solutions:}

These properties of multicast lead to security issues and vulnerabilities because of two reasons: the issues are multicast- specific or the issues also exist in unicast, but the unicast solutions do not apply.the three multicast properties leads to vulnerabilities and the areas of research that provide solutions to these issues. The multicast model delivers any traffic sent to the multicast address to the entire group. This means that any host can send data to the multicast group. This leads to two problems. First, group members need to be able to verify that messages received are from the intended source. Multicast source authentication solutions have been proposed to provide this functionality. Second, there should be mechanisms to restrict unauthorized sources from sending data to multicast groups due to the potential for denial-of-service attacks. Multicast sender access control solutions are necessary to defend against this threat.

\section{COMPARISON AND PERFORMANCE EVALUATION}

In this section, we evaluate MABS performance in terms of resilience to packet loss, efficiency, authentication latency and DoS resilience. As we discussed before, MABS does not assume any particular underlying signature algorithm. This is also true for all the literature multicast authentication schemes referenced in this paper. Therefore, all the discussions and evaluations of MABS and the literature works are under the assumption that they are using the same underlying signature algorithm. : We consider the authentication of digital streams over a lossy network. The overall approach taken is graph-based, as this yields simple methods for controlling overhead, delay, and the ability to authenticate, while serving to unify many previously known hash- and MAC-based techniques. The loss pattern of the network is defined probabilistically, allowing both bursty and random packet loss to be modeled.

The main challenges are fourfold. First, authenticity must be guaranteed even when only the sender of the data is trusted. Second, the scheme needs to scale to potentially millions of receivers. Third, streamed media distribution can have high packet loss. Finally, the system needs to be efficient to support fast packet rates. We use simulations to evaluate the resilience to packet loss. The metric here is the verification rate, i.e., the ratio of the number of authenticated packets to the number of received packets we compare MABS with some well-known loss tolerant schemes EMSS,augmented chain,PiggyBack,tree chain. These schemes are representatives of graph chaining, tree chaining, and erasure coding schemes and are widely usedin performance evaluation in the literature.  For EMSS  we choose the chain configuration , which has the best performance among all the configurations of length . For AugChain , we choose chain configuration. For PiggyBack , we choose two class priorities. For Tree chain [1], we choose binary tree. For all these schemes, we choose the block size of 256 packets and simulate over 100 blocks. We consider the random loss and the burst loss with a maximum loss length of 10 packets.

The verification rates under different loss rates are given in below Figures. OurMABS and Tree schemes have perfect resilience to packet loss in the sense that all the received packets can be authenticated. This is because all the packets in MABS and Tree schemes are independent from each other.

\begin{figure}
\centering
\resizebox{8cm}{6cm}{\includegraphics{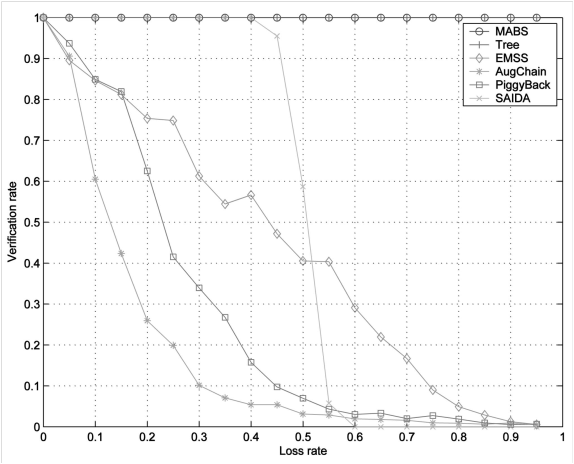}}
\caption{Verification rate under the random loss model}
\end{figure}

\begin{figure}
\centering
\resizebox{8cm}{6cm}{\includegraphics{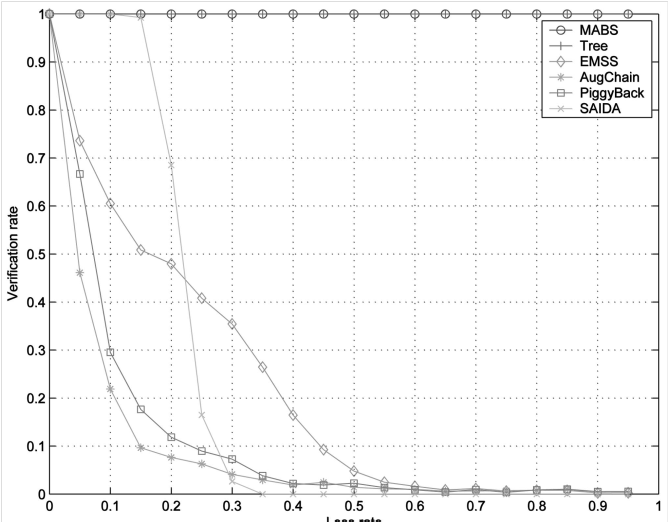}}
\caption{Verification rate under the burst loss model with the maximum burst length 10}
\end{figure}

\subsection{Comparison}

We consider latency, computation, and communication overhead for efficiency evaluation under lossy channels and DoS channels.

The notations used here are defined in Table :

\renewcommand{\labelenumi}{(\roman{enumi})}
\begin{enumerate}
\item  All the evaluations are carried out over on packets .
\end{enumerate}

\begin{figure}
\centering
\resizebox{8cm}{6cm}{\includegraphics{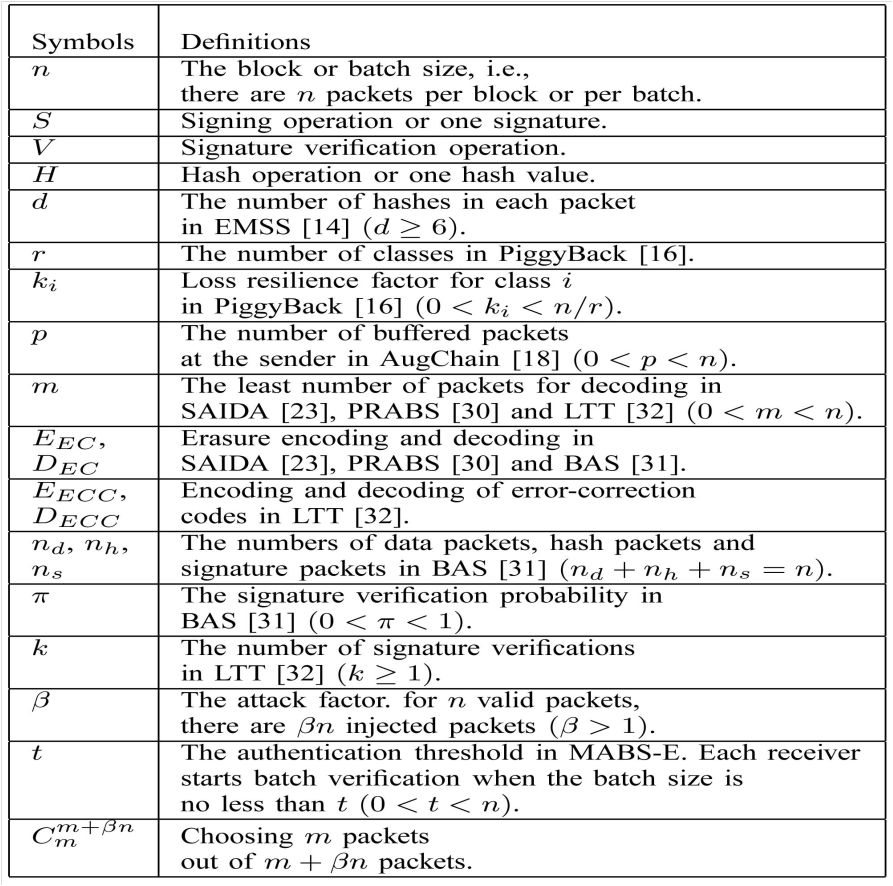}}
\caption{All the evaluations are carried out over n packets}
\end{figure}

\subsection{Resilience to Packet Loss:}

We use simulations to evaluate the resilience to packet loss. The metric here is the verification rate, i.e., the ratio of the number of authenticated packets to the number of received packets. We compare MABS with some well-known loss tolerant schemes EMSS augmented chain (AugChain) Piggyback tree chain (Tree) and SAIDA [2]. These schemes are representatives of graph chaining, tree Chaining and erasure coding schemes and are widely used in performance evaluation in the literature.

\subsection{Efficiency:}

We consider authentication latency, computation, and communication overhead for efficiency evaluation under lossy channels and DoS channels.

\subsection{Authentication Latency:}

The block-based approach requires each receiver to collect an entire block before authenticating every packet in the block. A larger block size achieves higher computation efficiency, but also incurs longer authentication latency. Our design does not have authentication latency .Because there is no relationship among packets and no limit on the number of packets in batch verification, each receiver can perform the batch verification over its buffered packets whenever higher-layer application require.

\subsection{DoS Resilience:}

DoS is a method for an attacker to deplete the resources of a receiver. processing forged packets from the attacker always consumes a certain amount of resources. The block-based approach has poor resilience to DoS. Because there is no filtering, each receiver has to recover the relationship among authentic packets mixed with forged packets. By using Merkle tree in our design, authentic packets and forged packets are separated into disjoint sets. Batch verification is carried out over each set. Therefore, each batch verification can authenticate a set of packets.

\subsection{Comparisons of Signature Schemes:}

We compare the computation overhead of three batch signature schemes in below table  RSA and BLS require one modular exponentiation at the sender and DSA requires two modular multiplications when r value is computed offline. Usually one c-bit modular exponentiation is equivalent to 1.5c modular multiplications over the same field. Moreover, a c-bit modular exponentiation in DLP is equivalent to a c/6-bit modular exponentiation in BLS for the same security level. Therefore, we can estimate that the computation overhead of one 1,024-bit RSA signing operation is roughly equivalent to that of 768 DSA signing operations (1,536 modular multiplications) and that of 6 BLS signing operations (each one is corresponding to255 modular multiplications).

\begin{figure}
\centering
\resizebox{8cm}{6cm}{\includegraphics{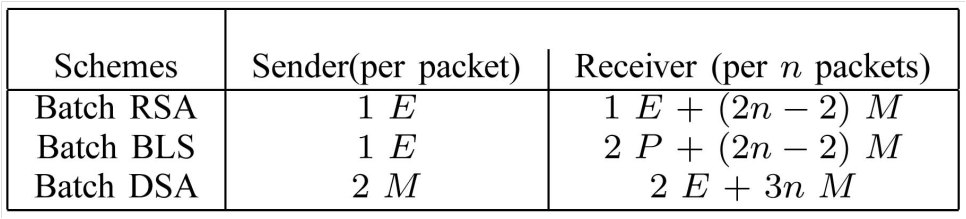}}
\caption{Computational Overhead of Different Batch Schemes}
\end{figure}

Below Table shows the comparisons between MABS-B and well known loss-tolerant schemes tree chain (Tree) ,EMSS,PiggyBack ,augmented chain (AugChain) , and SAIDA . MABS-B eliminates the correlation among packets. Each packet is independently sent out at the sender.

\begin{figure}
\centering
\resizebox{8cm}{6cm}{\includegraphics{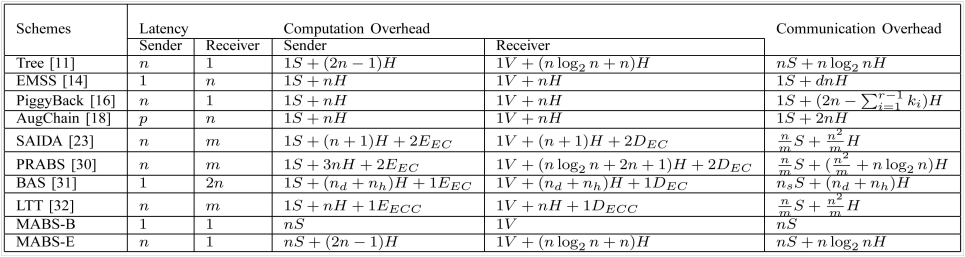}}
\caption{Comparisons over Lossy Channels:}
\end{figure}

We also compare the length of two popular hash algorithm MD5 [4] and SHA-1 [5] and the Signature length of three signature algorithms in MABS network generates a 320-bit signature. It is clear that by using BLS or DSA, MABS can achieve more bandwidth efficiency than using RSA, and could be even more efficient than conventional schemes using a large number of hashes.

\begin{table}
\renewcommand{\baselinestretch}{1}
\caption{Given the same security level as 1,024-bit RSA, BLS generates a 171- bit signature and DSA}
\begin{small}
\begin{center}
\begin{tabular}{|r|r|} \hline
       Schemes& Length(bits)  \\
    & \\ \hline
 $\rightarrow $ MD-5   & 125\\ \hline
 $\rightarrow$ SHA-1   &160 \\ \hline
 $\rightarrow$ RSA & 1024 \\ \hline
 $\rightarrow$BLS    & 171\\ \hline
 $\rightarrow$DSA    & 320\\ \hline
\end{tabular}
\end{center}
\end{small}
\end{table}

\section {CONCLUSIONS}
To reduce the signature verification overheads in the secure multimedia multicasting, block-based authentication schemes have been proposed. Unfortunately, most previous schemes have many problems such as vulnerability to packet loss and lack of resilience to denial of service (DoS) attack. To overcome these problems, we develop a novel authentication scheme MABS. We have demonstrated that MABS is perfectly resilient to packet loss due to the elimination of the correlation among packets and can effectively deal with DoS attack. Moreover, we also show that the use of batch signature can achieve the efficiency less than or comparable with the conventional schemes. Finally, we further develop two new batch signature schemes based on BLS and DSA, which are more efficient than the batch RSA signature scheme.

\section{The References Section}\label{references}

\noindent{\includegraphics[width=1in,height=1.7in,clip,keepaspectratio]{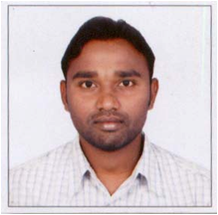}}
\begin{minipage}[b][1in][c]{1.8in}

{\centering{\bf {Srikanth Bethu}} is currently the Assistant Professor, Holy Mary Institute of Technology and Science, JNTU Hyderabad, Hyderabad. He obtained his Bachelor of Engineering from JNTU Hyderabad. He rece-}\\\\
\end{minipage}
ived his Masters degree in Computer Science and Engineering from Osmania University, Hyderabad. \\\\

\noindent{\includegraphics[width=1in,height=1.7in,clip,keepaspectratio]{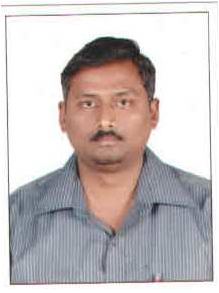}}
\begin{minipage}[b][1in][c]{1.8in}
{\centering{\bf{K Kanthi Kumar }}is a Associate Professor,Holy Mary Institute of Technology and Science, JNTU Hyderabad, Hyderabad. He was a Professor since 2010 with the Electronics and Communications Engineering,HITS college,JNTU Hyderabad.}\\\\
\end{minipage}
During the past 10 years of his service at various institutions he has over 5 research publications in refereed International Journals and Conference Proceedings..\\\\

\noindent{\includegraphics[width=1in,height=1.7in,clip,keepaspectratio]{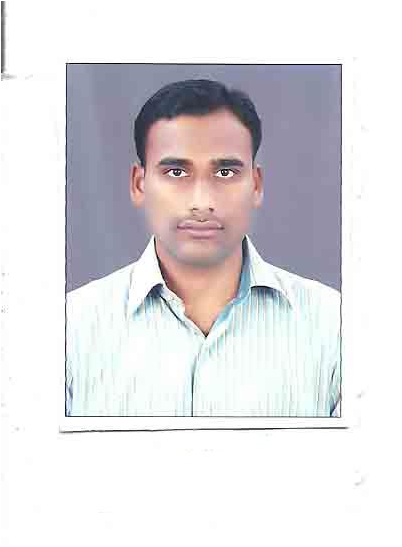}}
\begin{minipage}[b][1in][c]{1.8in}
{\centering{\bf{MD Asrar Ahmed }}is a Software engineer,Infosys,Hyderabad,since 2011.}\\\\
\end{minipage}

\noindent{\includegraphics[width=1in,height=1.7in,clip,keepaspectratio]{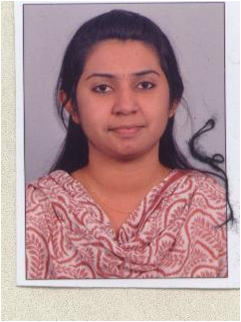}}
\begin{minipage}[b][1in][c]{1.8in}
{\centering{\bf{S.Soujanya }}is a Assistant Professor,Holy Mary Institute of Technology and Science, JNTU Hyderabad, Hyderabad.}\\\\
\end{minipage}
\end{document}